\documentclass{article}
\usepackage{hyperref}
\title{Do Black Holes Produce Phantom Energy?}
\author{George Svetlichny\footnote{Departamento de Matem\'atica, Pontif\'{\i}cia Universidade Cat\'olica, Rio de Janeiro, Brazil \newline
svetlich@mat.puc-rio.br \hfill \url{http://www.mat.puc-rio.br/\~svetlich}}}
\begin{document}
\maketitle
\begin{abstract}
We conjecture that black holes, whether accreting matter  or not, can also lose  mass to phantom energy production. Such a process can be made plausible in theories that allow for non-local connections between short and long wavelength modes. If true, observational data should show an exchange of energy from matter to phantom at some cosmic epoch, and we discuss a simple model.  Some remarks on an appropriate quantum gravity theory and related issues are also presented.
\end{abstract}

\section{Introduction} Under the usual energy conditions assumed in general relativity, space-time singularities are generic in a wide range of situations \cite{hawpen}. In quantum gravity one expects that what would be a classical singularity would be replaced by something of normal status within the theory. The prevailing view seems to be that the singularities would be ``smeared-out" by quantum behavior and so in essence disappear and lose their problematic nature, but such passive amelioration of the problem is not the only possibility. 
The exact quantum nature of a would-be classical singularity must await a detailed true quantum gravity, but one can speculate as to what the semi-classical space-time description would be around such regions. How is Einstein's theory modified due to quantum gravity's reaction to what classically would be the formation of singularities?  Schuller and Wohlfarth \cite{schu-wohl076} have propose a semi-classical theory based on the idea that quantum gravity would impose upper and lower bounds on the sectional curvatures, introducing thus two (related) length scales, however we shall not make use of their assumptions. 

Quantum field theory allows for regions that violate the energy conditions, and thus diminish the tendency to form singularities. However, according to quantum inequalities \cite{fewst} these violations are limited, and, for instance, do not seem to allow for useful wormhole stabilization \cite{rom}. Onemli and Woodard \cite{on-wo}  have even  argued that such violations can happen on a cosmic scale, but again they are greatly limited.  Quantum gravity however could be more effective and it would be natural to suppose that its reaction  to a potential singularity is to create regions where according to a semi-classical description some energy conditions are violated. 

The only  empirical evidence of possible energy-condition violations not subject to quantum inequalities comes from the accelerated expansion of the universe. The current observations are consistent with a dark energy component in which the equation of state \(p=w\rho\) is consistent with \(w\approx -1\) and in fact favors \(w<-1\) \cite{ma-sa-am}.  An energy component  with \(w<-1\) is dubbed ``phantom energy" and violates the dominant energy condition.
One could now  posit that the formation of some types of would-be singularities causes quantum gravity to create phantom energy which then impedes the formation of a true singularity and ``heals" the problematic region of space-time. If we are to associate phantom energy with would-be singularities, then the only candidates for these at the present epoch of the universe is the population of black holes, and so we conjecture that the phantom energy we see is due to these black holes.

In the rest of this introduction we argue that such an unconventional idea can be made plausible in theories that allow non-local connections  between short and long wavelength modes. 
Section \ref{section.simplemodel} presents a very simplified model of  exchange between matter and phantom energy, deriving verifiable consequences if such black-hole mediation is true. In section \ref{section.QG} we describe some basic characteristics of a putative  quantum gravity theory incorporating short-long wavelength connections. Section \ref{section.causality.ontology}  takes up issues of ontology and causality, and Section \ref{section.conclusions} some concluding considerations.

Specifically we posit that  processes at the Planck scale near what would be a classical singularity can non-locally influence long-wavelength processes (which in our universe could be on the Hubble scale) not impeded by event horizons nor by the usual causality relations (more on this in Sections \ref{section.QG} and \ref{section.causality.ontology} where we argue there are no true causality violations). We come to this hypothesis from considerations that at the Planck scale ordinary quantum mechanics may no longer be the true mechanics and that some sort of modified mechanics is operant. A form of modified quantum mechanics known as ``non-linear quantum mechanics" has been discussed in the literature for some time (see \cite{nlqmatplanck} for a brief survey) but for our purposes here, only the cited assumptions are relevant and other frameworks can also lead to the same conclusions. Thus we allude to non-linear quantum mechanics mainly as a source of our ideas, not necessarily as an argument in favor of this specific theoretical framework. We shall refer to the quantum gravity here contemplated as ``non-local quantum gravity", leaving linearity (or lack thereof) as a separate issue, and using ``nonlinear" in situations when that was the specific assumption.

We argued in \cite{amply} that a nonlocal short-long wavelength connection is present in the nonlinear Schr\"odinger equation of Doebner and Goldin \cite{dg} which is related to the diffeomorphism group, suggesting hence a short-long connection in quantum geometry. This is reminiscent of the so called IR/UV entanglement in non-commutative field theories \cite{do-ne}. Non-linear quantum gravity and non-commutative space-time theories may well share many common features, among them modified dispersion relations for ultra-high energy particle propagation \cite{svetcr}. 

An immediate apparent difficulty with assuming Planck-scale nonlinearities is that nonlinear quantum theories  are generally singular for highly localized states. This apparently exacerbates the singularity issue. One needs truly new physics to overcome this, but, it is precisely this singular behavior, along with other properties, that allows for such a truly new phenomenon, the possibility of nonlocal short-long wavelength connection, maintaining a  suppression of the problematic nonlocal effects at low energies \cite{amply}.

The dominant energy condition is necessary to prove the black-hole area theorem \cite{hawpen}, stating that black-hole horizon areas increase. A violation of the dominant energy condition, as happens with phantom energy, would allow for a decrease of the horizon area. Babichev, Dokuchaev and Eroshenko \cite{babi.etal089} showed that in the presence of phantom energy black holes can lose their mass. They made no assumptions concerning the source of the phantom energy. We here suggest that it may be connected to the black hole itself. It would be a nonlocal quantum gravity reaction to the would-be classical singularity within the horizon. Quantum gravity would thus, by Planck-scale processes near the would-be singularity excite long-wavelength modes of phantom energy overcoming the horizon restriction through non-locality, causing phantom energy to appear part of which is then accreted by the black hole diminishing its mass  (disregarding other inflow of matter) and in the long run eliminating the would-be singularity. One way of envisioning this process is through a typical quantum metaphor: The vacuum contains virtual states that entangle short and long wavelengths. The would-be singularity in a black hole acts as a position measuring apparatus for Planck-scale states, which are somehow incorporated into the would-be singularity.  The corresponding  long-wavelength partners are then projected into definite states of phantom energy, which is now free to interact with the black hole to diminish its mass. We must warn the reader though that this is merely a metaphor, much like the metaphor explaining Hawking radiation through one of the particles of a virtual particle pair passing through the horizon and the other escaping.  

Aside from the primordial ones, black holes in our universe are formed by mass accretion which continues for a very long time (billions of years). This allows black holes to be recycling stations, converting the accreting ordinary and dark matter into phantom energy. As the mass source gets depleted the phantom energy then eventually extinguishes the black hole leaving behind a possible remnant of matter, not massive enough to form a black hole, along with some remaining phantom energy. 

The remaining phantom energy is problematic as it would generally lead to a Big Rip singularity \cite{ca-ka-we}. If however we feel phantom energy is singularity related, then it's natural to suppose that the inflationary period of the universe was related to the apparent initial classical singularity and that the inflation energy was phantom. The universe avoided a Big Rip back then by generating matter, and it's valid to suppose this would be true of the phantom energy we see today. In the end, phantom energy may be a transitory object, used by quantum gravity to resolve some singularities and eliminated by quantum gravity by transforming it to ordinary matter to eliminate the Big Rip singularity. In sum: {\sl phantom energy is quantum gravity's answer to the Crunch, matter is it's answer to the Rip.} One can look upon matter genesis as a short-long wavelength connection in the other direction. The stretched-out modes of phantom energy disappear in favor of localized matter. Ironically enough, the final role of black holes may be to counteract gravity's tendency to clump matter by spreading it out again.

\section{A simple model}\label{section.simplemodel}

Consider a flat FRW cosmology with  metric
\[ds^2=dt^2-a(t)^2(dx^2+dy^2+dz^2).\]
We assume a two sector cosmology, dark matter and phantom energy. We assume dark matter to be pressureless, which
is corroborated by observational bounds on its equation of state \cite{mull}. We also posit a black hole population of density \(\mu(t)\) per co-moving volume, assuming it contributes negligibly to the total energy density of the universe (at the present epoch, baryonic matter contributes about \(0.4\%\) \cite{benetal}).
 As they accrete dark matter the black holes  produce phantom energy with equation of state \(p_{\rm ph}=w\rho_{\rm ph}\) with \(w<-1\). Thinking of the dark matter as dust particles, the presence of the black holes gives each such particle a certain probability per unit time to be converted to phantom energy. We can thus posit the following evolution equation for the density \(\rho_m\) of dark matter:
\begin{equation}\label{dmd}
\frac{\frac d{dt}(a^3\rho_m)}{a^3\rho_m}=-\sigma(t)\mu(t)=-b(t).
\end{equation}

Here \(\sigma(t)\) is the probability per unit time that a black hole would convert a unit of dark matter into phantom energy. We denote the product \(\sigma \mu\) by \(b\). This latter function contains a multitude of sins: the black hole population density, the details of the accretion process, the yet unknown parameters of the conversion process, the dark matter distribution and dynamics, etc., just to mention a few. The equation thus reflects very little of the actual hypothesis of the role of black holes as recycling stations, but one can still draw some conclusions. 

We assume that the total density \(\rho=\rho_m+\rho_{\rm ph}\) is conserved. Due to the postulated nonlocal nature of the conversion process this may not be strictly true on the local scale (as discussed in Section \ref{section.QG}), but it's natural to assume it on a large scale and a conservation law can be taken as a good approximation. The dynamics of the universe can be thus modelled by a version of the energy exchange equations as in \cite{ma-sa-am} except for the direction of the transition which is now from dark matter to phantom. We thus come to the FRW equations:

\begin{eqnarray} \label{fried}
\left(\frac{\dot a}a\right)^2 &=& \frac{\kappa_0^2}3(\rho_m+\rho_{\rm ph}),\\ \label{darkcont}
\dot\rho_m +3H\rho_m &=& - b\rho_m,\\ \label{phantomcont}
\dot\rho_{\rm ph} +3\gamma H\rho_{\rm ph}&=&  b\rho_m.
\end{eqnarray}
here \(\kappa_0=8\pi G\), \(H=\dot a/a\) is the Hubble parameter, and \(\gamma=w+1<0\).

At this point we can think of \(b\) as merely a phenomenological function appearing in (\ref{darkcont}-\ref{phantomcont}) whose existence is guaranteed by conservation of total density. 
The equations would only be valid during a certain time period of cosmic evolution, which should include the current epoch.  Prior and subsequent to this epoch other dynamics would dominate the universe.

It is convenient to rewrite equations (\ref{fried}-\ref{phantomcont}) in terms of the quantities \(a\), \(\rho\), and \(\Omega_m=\rho_m/\rho\). We have

\begin{eqnarray}\label{fried2}
\left(\frac{\dot a}a\right)^2 &=& \frac{\kappa_0^2}3\rho,\\\label{denscont}
\dot\rho+3(\gamma -w\Omega_m)H\rho&=&0, \\ \label{omegadot}
\dot\Omega_m +(b-3wH)\Omega_m &=& -3wH\Omega_m^2.
\end{eqnarray}

Now equation (\ref{denscont}) does not involve \(b\) and is a consequence of the continuity equation and the hypothesis of two dark sectors (matter and energy) with their respective equations of state; it is essentially a kinematic relation. Any dynamical information has to be gathered from equation (\ref{omegadot}). Knowing \(a(t)\)
one can solve for \(H\), \(\rho\), \(\Omega_m\), and \(b\). We have:
\begin{equation}\label{omegam}
\Omega_m = 1+\frac1w+\frac2{3w}\frac{\dot H}{H^2}=1+\frac1{3w}-\frac{2q}{3w},
\end{equation}
\begin{equation}\label{b}
b=  3wH(1-\Omega_m)-\frac{\dot \Omega_m}{\Omega_m},
\end{equation}
where in (\ref{omegam})  we've introduce the deceleration parameter \(q=-(\ddot a/a)/H^2\). The right hand side of (\ref{omegam}-\ref{b}) can in principle be evaluated by empirical data. Note that \(\Omega_m\) is dependent of the second derivative \(\ddot a\), the cosmic acceleration, while \(b\) is dependent on the third derivative, the cosmic jerk.
Taking a present estimated value \(q_0=-.55\) \cite{vireetal}  and \(w=-1.05\) \cite{knop}, we calculate \(\Omega_m=0.3 \) which is consistent with the currently accepted value \(0.27\pm 0.04\) \cite{benetal}. All this means is that the two sector model of pressureless dark matter and dark energy with equation of state close to a cosmological constant is a good description of the present universe. 

The only definite prediction we can make about the \(b\) function is that it must be positive as some time. It is positive if energy flows from matter to phantom, and negative if in the other direction.  The condition \(b\ge 0\) is
\begin{equation}
\frac{\dot \Omega_m}{\Omega_m}\le 3wH(1-\Omega_m).
\end{equation}
This is a relation that can be tested empirically. 
Using presently accepted values \cite{benetal} of \(\Omega_{m0}=0.27\) and  \(H_0=71 \,{\rm km\,s^{-1}/Mpc}\), with \(w_0=-1.05\) \cite{knop}  the inequality yields 
\begin{equation}\label{dotomeganow}
\dot \Omega_{m0} \le -0.045\,{\rm Gyr}^{-1}. 
\end{equation}

If black-hole mediation of phantom energy production is true, the sign of the right hand side of (\ref{b}) should be positive during some part of the universe's evolution after the radiation dominated era. Whether it should still be positive at the present coincidental moment when dark matter and dark energy have densities of the same order of magnitude (the turn-around point) is not a-priori clear as phantom energy could also be transferring energy to dark matter by some mechanism making the net energy flow from phantom to matter. 
If \(b\) is positive at some epoch, then at least the conjecture that energy flows from matter to phantom is not rejected, even though a connection to black holes does not follow logically.  If found false  for the whole relevant epoch, the present theory fails.  Such a failure will impose non-trivial constraints on any nonlinear version of quantum gravity eliminating some of the more striking aspects the theory might have otherwise. In any case, the right-hand side of (\ref{b}), or its more sophisticated variant, is an important empirical datum in our search for a better understanding of quantum features of space-time.

Though there is data concerning the evolution of dark matter \cite{bacoetal}, it doesn't seem to be sufficiently precise to determine the derivative \(\dot \Omega_m\) with any accuracy. Szyd{\l}owski \cite{szyd} argues for a negative sign for \(b_0\) based on SNIa observations, but again, given the uncertainties in the data, a positive sign cannot be entirely ruled out. 

To go beyond the analysis made above and present a model for the \(b\) function, one must have some idea of the conversion process. This faces a series of challenges. In the first place, 
it's a hard task to even approximate what would be the actual dynamics of a black hole converting itself to phantom energy, even disregarding any other in-falling matter. The calculations of  Babichev, Dokuchaev and Eroshenko \cite{babi.etal089} are based on those of Michel \cite{mich} who neglects the effect of the accreting matter on the metric, which is taken to be a static Schwarzschild black hole. Thus it is valid only for a low density gas accreting onto a very massive black hole. Though the sign of the mass change (negative for phantom energy) is reliably predicted by the authors of \cite{babi.etal089} their analysis of the actual temporal evolution must be questioned. It is even possible to have a {\em  static\/} black hole surrounded by phantom energy. Kiselev \cite{kise} presents exact static black hole solutions in the presence of quintessence (\(p=w\rho\) with \(-1<w<-\frac13\)), but these solutions make perfect sense for \(w<-1\). This again shows that the exact dynamics of black holes in the presence of phantom energy is still to be worked out. Another complicating factor for the present theory is that since phantom energy is to be produced in a non-local way the usual conservation equation \(T^{\mu\nu}{}_{;\nu}=0\) for the {\em  true\/} stress-energy tensor cannot hold, and this is the basis for the usual description of spherical accretion. Einstein's equation \(G_{\mu\nu}=8\pi GT_{\mu\nu}\) also cannot hold as the left-hand side automatically satisfies the conservation equation for purely geometric reasons.  For the actual semi-classical dynamics of the conversion of black-hole mass to phantom energy we have to seek another description. Continuous creation theories such as that of Hoyle and Narlikar \cite{hoyl-narli} may be of help, although they are rather ad hoc and are tailored for different situations.

One immediate apparent difficulty with  black-hole mediated phantom energy production is the delay between the formation of black holes and the rise of phantom energy density. The present acceleration of the universe started at some red-shift parameter \(z<1\) and likely near \(z=0.5\) \cite{riess}. Models of black hole population evolution show a peak in the rate around \(z=2.5\) and black holes were forming  at much higher red-shifts \cite{ma-de-si}. There could be many reasons for the delay. If the rate of phantom energy production is low  enough then, in spite its tendency to grow at an accelerated rate, its effect would only manifest itself at much later times. The universe was likely decelerating for \(z>0.5\) \cite{riess, tu-ri} and phantom energy before this had to work against the deceleration. Phantom energy may be converting to matter by some mechanism, delaying its appearance in a dominant form. More importantly though, phantom energy production is to be associated to events near the would-be singularities of the black holes, but these events do not have any intrinsic relation to the cosmic proper time of the surrounding universe. Hence the ``onset" of phantom energy production cannot, under present conceptions, be clearly related to any value of the cosmic time. In fact the only covariant way of looking at this is to assume phantom energy was present all the time (see Section \ref{section.QG}). The empirical delay should be looked upon as an important clue toward a final  quantum gravity theory rather than as falsifying evidence.

\section{Non-local Quantum Gravity}\label{section.QG}
To give some idea of our view of quantum gravity, a metaphorical description will be useful:  We assume space time can be described by a very large number of degrees of freedom.  At Planck energies space-time undergoes something like a phase-transition with the Planck temperature ($\approx 1.417 \times 10^{32}\, {\rm K} $ ) serving as the critical temperature. Space-time is then thought of as akin to a ferromagnet with the space-time metric (and possibly other physical fields) being analogous to  magnetization. Above the critical temperature it becomes ``demetrized" (analogous to being demagnetized). The degrees of freedom in the disordered phase no longer have a strict metric relation (since order is precisely ``metrization") to the ones in the ordered phase and so can ``reenter" space-time in a non-local way giving rise  to phantom energy, and possibly other manifestations. This view differs from the more conventional views in its vision of would-be singularities, which are seen now as loci of short-long wavelength non-local connections. Ideas reminiscent of some of these have already been voiced in the literature. Markopoulou and Smolin \cite{tati} present a model in which quantum behavior can be derived from stochastic motion of degrees of freedom whose relative metric relations are different from other non-metric ones (adjacency in a graph), and so those metrically distant may have  close intrinsic relations (and vice-versa), providing a form of non-locality.  These authors did not consider degrees of freedom not metrically related to others, but their model can be easily modified to do so. Jacobson \cite{jacob} derives Einstein's equation from an appropriate thermodynamic equilibrium assumption, and speculates that at high energies one enters a nonequilibrium regime suggesting thus a thermodynamic view of quantum gravity.

Non-local quantum gravity should possess at least one new characteristic physical parameter \(\nu\) which on the one hand can be interpreted as the suppression factor of non-local effects at low-energies, or alternatively combined with the Planck length \(L_p\approx  10^{-33}\,{\rm cm}\) as \(R=L_p/\sqrt{\nu}\), determine the scale of long-wavelength modes  which Planck-scale processes can excite non-locally \cite{amply}. Whether \(\nu\) should be constant or vary with cosmic time is at present not clear.

Kowalski-Glikman and Smolin \cite{KG.Smo} consider a new length scale which they estimate to be about \(10^{60}L_p\), about a thousandth of  the present Hubble radius. They base it on the disparity between the observed dark energy density and that predicted by quantum field theory. In our view the present dark energy density does not correspond to any fundamental physical parameter,  just a contingent result of previous black hole formation.

't~Hooft \cite{thoo} presented a wonderfully picturesque description of a spherical collapse to a black hole of a large shell made up of working television sets separated by large distances. At some instant, still with considerable distance between the sets, a horizon forms and the shell, in its proper time, enters it. Under conventional general relativity what a distant observer will see (neglecting Hawking radiation) is an asymptotic approach of the shell to the horizon surface, never crossing it from her point of view. Under our quantum gravity proposal the distant observer continues to see the shell collapse as the space around her begins to fill with phantom energy. Eventually the matter reaches high density and temperature. The final remnant of the collapse would be a sphere of ordinary matter (a star maybe) not massive enough to form a black hole. There would be no horizon nor singularity and a certain portion of the initial matter will have been converted to phantom energy.

Consider the same situation now from the point of view of a Cauchy surface containing the initial shell of television sets. From our perspective the Cauchy surface has to also contain a description of phantom energy. The state of this phantom energy cannot be given arbitrarily as it must in the end have the necessary form to eliminate the would-be singularity that conventional general relativity predicts. Thus for the phantom energy one must impose some {\em  future\/} boundary conditions. This is what corresponds to the  non-locality  of non-relativistic non-linear quantum mechanics. Future boundary conditions in quantum gravity have been considered in other contexts \cite{ge-ma-ha}. Of particular interest is the proposal of G.~T.~Horowitz and J.~Maldacena \cite{homa} in which a unique final state in the interior of a black hole is invoked to solve the black hole information loss problem. Thus, as in our view, the  would-be singularity is given a special quantum role. In this regard Yurtsever and Hockney \cite{yurt-hock} have even argued that quantum mechanics becomes non-linear to accommodate such an unique final state. See though a contrary consideration by Ge and Shen \cite{ge-she}.

We would maintain that information gets preserved in the final collapse remnant of ordinary matter along with the generated phantom energy.

Phantom energy cannot be subject {\em  only\/} to future boundary conditions. Presumably the remnant phantom energy after the would-be black hole singularity is eliminated would be subject to some evolution dynamics of its own leading to a would-be Big Rip which is then eliminated by another action of quantum gravity generating matter, in sum: {\em  matter happens.\/} 

Short and long wavelengths are intrinsically different.\footnote{\(T\)-duality \cite{polc} in string theory may be brought forth as a counterargument, but string theory relies on unmodified quantum mechanics at all scales, which we like to doubt.}
This asymmetry is reflected in the temporal asymmetry for phantom-energy boundary conditions, the choice of a {\em  future\/} one for black hole singularity prevention. The cosmological arrow of time may in the end be the result of this asymmetry. The Doebner-Goldin equation is not time-reversal invariant and we consider this as the non-relativistic remnant of what would be a fundamental asymmetry in a nonlinear quantum gravity theory. For the case of avoidance of the Big Rip singularity it does not seem necessary to impose future boundary conditions, which point out again the fundamental asymmetry.

\section{Ontology and Causality }\label{section.causality.ontology}
Non-locality always raises questions of ontology and causality. Quantum gravity raises new interpretational issues. We address one stemming from the above proposal.

Consider a universe with no black holes and no phantom energy. I am faced with the decision of whether to build a 't~Hooft sphere or not. If I do so, then after the conversion process, when my proper time is \(T\),  I'm left with an ordinary matter remnant and phantom energy. After \(T\) my description   of space-time must include some phantom energy even prior to my decision. If I do not build the sphere then after \(T\) my description would not include phantom energy at any time in the universe. This seems superficially  an influence on the past, my decision to build the sphere or not, determines if there is phantom energy or not {\em  before\/} my decision. This is no more paradoxical than any other delayed choice quantum experiment \cite{whe}, in fact as the creation of phantom energy is a quantum effect, this situation is {\em  exactly\/} a delayed choice experiment. The two alternative futures are incompatible and so there is no real contradiction in their descriptions of space-time. The proper ontology for this, to be elaborated in a future publication,  is that {\em  space-time is an organizing principle for actuality\/}.  In the quantum gravity universe, my past, the content of the interior of my past light cone, (whatever its ontological status might be) would not be an immutable ``has been". My description of it would change with different actualities. In the two alternative actualities mentioned above, even in the overlapping parts of the past of my two possible futures, I must in one include phantom energy and exclude it in the other. Phantom energy is quantum mechanics writ large, the short-long wavelength connection forces some large-scale features of the the universe to behave in a quantum mechanical way. The idea that cosmic-scale phenomena may have quantum nature was also stated by Sorkin \cite{sork009} in connection with causal set theory. The possible quantum nature of cosmic-scale phantom energy excitations could in the end make any quasi-classical approximation to such a space-time problematic. In a sense, as space-time expands under phantom energy, it may be becoming more and more quantum.

Causality problems can only be addressed once an explicit mathematical theory is available. The burden of proof in these respects is on one who claims there is a violation, since the one who defends causality has no way of envisioning all possible mechanisms by which it can fail. His a priori defence can only be of a straw-man kind. We shall knock down two obvious straw men. To do more one would need a  mathematical theory.

First straw man:  My colleague claims  she has a phantom energy detector, and so claims that she can tell if I decide to build the 't~Hooft sphere or not, and doing so cajole me to the opposite decision thereby creating a time-loop paradox. However, the non-local connection is of the short-long wavelength type. For her detector to work its size must be of order  \(R\). In order to have a true detection event, information from all parts of the detector must reach a single point and this can only happen at time of order \(R\) after the detector has been engaged. By this time the decision would have been taken and no time loop is possible. The operant principle here is: {\em  information about the future has to be so spread out that by the time it is gathered together, the future has already happened.\/} The usual arguments for causality violation in non-linear quantum mechanics through the measurement process \cite{polcal} are based on formalism and  do not take into account the physical makeup of the apparatus, their sizes, and the fact that the experiments take place in an evolving universe. If one of the detectors is forced to be of cosmic size causality violation is avoided.

Second straw man: Assume for the sake of the argument that local phantom energy detection is possible. Assume I can create mini black holes at will and that phantom energy appears instantaneously in its rest frame and extends to a distance \(R\) which is comparable to the Hubble radius.  The phantom energy released by a black hole created by me is detected by my colleague far away who creates a black hole in a moving frame allowing me to detect her phantom energy prior to my decision to create the first black hole. This is another time-loop paradox. 
However, even if local phantom energy detectors exist, if \(R\) is of Hubble order, then the long-wavelength modes could, so to speak, ``feel the shape of the universe" and be exited accordingly. In our universe this would mean that these modes could react to the preferred cosmological reference frame. The phantom energy of a boosted black hole would not be correspondingly boosted, so that creation of phantom energy would not be locally Lorentz invariant, preventing again time-loop paradoxes.

\section{Conclusions}\label{section.conclusions}

Considerations of quantum nonlinearities at the Planck scale lead naturally, albeit by a thin logical thread, to consider an active role for would-be space-time singularities. They would, through non-local effects couple short and long wavelength processes, what could be called a Planck-Hubble connection. Such a connection has consequences independently of its postulation in non-linear theories. If true, all theories based on unmodified quantum mechanics being valid at all scales must be reconsidered. Being a radical alternative is not sufficient ground for consideration even if some  outstanding problems are better addressed in this way (see \cite{nlqmatplanck}). The present proposal has empirical consequences that can be tested and so, complying with a basic scientific requirement,  is falsifiable.  The loss of mass of black holes to phantom energy, besides giving a positive value to the empirically accessible right hand side of (\ref{b}), would, if the rate is sufficiently large, influence all other behavior of  these objects. If the theory is falsified,  the thin logical thread is in fact broken at some point. One way or another, the empirical data would furnish invaluable information for a better understanding of our universe. 
 
\subsection*{Acknowledgements} This research received partial financial support from the Conselho Nacional de Desenvolvimento Cient\'{\i}fico e Tecnol\'ogico (CNPq).


\begin{thebibliography}{cc}

\bibitem{hawpen}S.~W.~Hawking and G.~F.~R.~Ellis, ``The large-scale structure of space-time", Oxford University Press, 1973.

\bibitem{schu-wohl076} F.~P.~Schuller and M.~N.~R.~Wohlfarth, ``Classical limit of quantum gravity in an accelerating universe", gr-qc/0411076. 

\bibitem{fewst} C.~J.~Fewster, ``Energy Inequalities in Quantum Field Theory", math-ph/0501073.

\bibitem{rom}T.~Roman, ``Thoughts on Energy Conditions and Wormholes", to appear in the Proceedings of the Tenth Marcel Grossmann Meeting on General Relativity and Gravitation, gr-qc/0409090.


\bibitem{on-wo} V.~K.~Onemli and R.~P.~Woodard, {\sl Physical Review \/} \textbf{D70}, 107301 (2004), gr-qc/0406098.


\bibitem{ma-sa-am} E.~Majerotto, D.~Sapone and L.~Amendola, ``Supernovae type Ia data favour coupled phantom energy", astro-ph/0410543.

\bibitem{nlqmatplanck} G.~Svetlichny, ``Nonlinear Quantum Mechanics at the Planck Scale", quant-ph/0410230.


\bibitem{amply} G.~Svetlichny, ``Amplification of Nonlocal Effects in Nonlinear Quantum Mechanics by Extreme Localization", quant-ph/0410186.

\bibitem{dg} H.-D.~Doebner and  G.~A.~Goldin,  {\sl Physics Letters\/}  A~\textbf{162}, 397 (1992). 


\bibitem{do-ne}    M.~R.~Douglas  and   Nikita A.~Nekrasov, {\sl Review of Modern Physics\/} \textbf{73}, 977 (2001), hep-th/0106048.


\bibitem{svetcr}G.~Svetlichny, {\sl Foundations of Physics Letters} {\bf 17}, 197 (2004),  hep-th/0305100.

\bibitem{babi.etal089} E.~Babichev, V.~Dokuchaev and Yu.~Eroshenko,  {\sl Physical Review Letters\/} \textbf{93}, 021102 (2004), gr-qc/0402089.

\bibitem{ca-ka-we}    R.~R.~Caldwell, M.~Kamionkowski and N.~N.~Weinberg,  {\sl Physical Review Letters\/} \textbf{91}, 071301 (2003), astro-ph/0302506.


\bibitem{mull} C.~M.~Mueller, {\sl Physical Review\/} \textbf{D71},  047302 (2005), astro-ph/0410621.

\bibitem{benetal}C.~L.~Bennett, {\sl et al.}, {\sl The Astrophysical Journal Supplement} \textbf{148}, 1 (2003), astro-ph/0302207.

\bibitem{vireetal} J.-M.~Virey, P.~Taxil, A.~Tilquin, A.~Ealet, C.~Tao and D.~Fouchez, ``On the determination of the deceleration parameter from Supernovae data", astro-ph/0502163.

\bibitem{knop} R.~A.~Knop, {\sl et al.}, {\sl The Astrophysical Journal\/} \textbf{598}, 102 (2003), astro-ph/0309368.

\bibitem{bacoetal}D.~J.~Bacon {\sl et. al.}, ``Evolution of the Dark Matter Distribution with 3-D Weak Lensing", astro-ph/0403384.

\bibitem{szyd}M.~Szyd{\l}owski,``Cosmological Model with Energy Transfer", astro-ph/0502034.

\bibitem{mich} F.~C.~Michel, {\sl Astrophysics and Space Science\/} \textbf{15}, 153 (1972).


\bibitem{kise}V.~V.~Kiselev, {\sl Classical and Quantum Gravity\/} \textbf{20},  1187 (2003), gr-qc/0210040.


\bibitem{hoyl-narli}    F.~Hoyle, and  J.~V.~Narlikar,  {\sl Procedings of the Royal Society\/} \textbf{A290}, 162 (1966).

\bibitem{riess} A.~G.~Riess, et al.,  {\sl The Astrophysical Journal} \textbf{607}, 665 (2004), astro-ph/0402512. 
     

\bibitem{ma-de-si}A.~Mahmood, J.~E.~G.~Devriendt and J.~Silk, ``A simple model for the evolution of super-massive black holes and the quasar population", astro-ph/0401003.

\bibitem{tu-ri} M.~S.~Turner and A.~Riess, ``Do SNe Ia Provide Direct Evidence for Past Deceleration of the Universe?",  astro-ph/0106051.

\bibitem{tati} F.~Markopoulou and L.~Smolin, {\sl Physical Review\/} \textbf{D70}, 124029 (2004), gr-qc/0311059.

\bibitem{jacob} T.~Jacobson, {\sl Physical Review Letters\/} \textbf{75}, 1260 (1995), gr-qc/9504004.

\bibitem{KG.Smo} J.~Kowalski-Glikman and L.~Smolin,   {\sl Physical Review\/} \textbf{D70},  065020 (2004), hep-th/0406276.

\bibitem{thoo} G.~'t~Hooft, ``Horizons"
 Lecture presented at the Erice School for Sub-Nuclear Physics, Sept. 2003, gr-qc/0401027.
 
 \bibitem{ge-ma-ha}M.~Gell-Mann, J.~B.~Hartle, `` Time Symmetry and Asymmetry in Quantum Mechanics and Quantum Cosmology", in {\sl Physical Origins of Time Asymmetry\/}, edited by J.~Halliwell, J.~Perez-Mercader, and W.~Zurek, Cambridge University Press, Cambridge,(1994), gr-qc/9304023.  
 
\bibitem{homa}   G.~T.~Horowitz and J.~Maldacena, ``The Black Hole Final State", {\sl Journal of High Energy Physics} \textbf{0402}, 008 (2004), hep-th/0310281.

\bibitem{yurt-hock} U.~Yurtsever and G.~Hockney,  {\sl Classical and Quantum Gravity\/} \textbf{22},  295 (2005), gr-qc/0409112.
 
\bibitem{ge-she}Xian-Hui Ge and You-Gen Shen, ``Reconsidering the black hole final state in Dirac fields",  hep-th/0501131.

\bibitem{polc} J.~Pochinski, ``String Theory", 2 vols., Cambridge University Press, 1993.

\bibitem{whe}
J.~A.~Wheeler, in {\sl Mathematical Foundations of Quantum Theory\/}, edited by
A.~R.~Marlow, Academic Press, New York, 1978, p. 9; in {\sl Quantum Theory and
Measurement\/}, edited by J.~A.~Wheeler and W.~H.~Zurek, Princeton University
Press, 1983, p. 182;
R.~B.~Griffiths,  {\sl  Fortschritte der Physik\/} \textbf{46}, 741 (1998), quant-ph/9810016.

\bibitem{sork009} R.~D.~Sorkin, ``Causal Sets: Discrete Gravity", to appear in the proceedings of the Valdivia Summer School, edited by A.~Gomberoff and D.~Marolf, gr-qc/0309009.

\bibitem{polcal}N.~Gisin,   {\sl Helvetica Physica Acta\/}  \textbf{62}, 363 (1989);  M.~Czachor, {\sl Foundations of Physics Letters} \textbf{4}, 351 (1991);  J.~Polchinski,  
{\sl Physical Review Letters\/} \textbf{66}, 397 (1991);  G.~Svetlichny,
 {\sl Foundations of Physics} \textbf{28}, 131 (1998), quant-ph/9511002;  W.~Luecke,
``Nonlocality in Nonlinear Quantum Mechanics",
quant-ph/9904016.
\end{thebibliography}
\end{document}